# Enhancement mode double top gated MOS nanostructures with tunable lateral geometry

E.P. Nordberg,[1,2] G.A. Ten Eyck,[1] H.L. Stalford,[1,3] R.P. Muller,[1] R.W. Young,[1] K. Eng,[1] L.A. Tracy,[1] K.D. Childs,[1] J.R. Wendt,[1] R.K. Grubbs,[1] J. Stevens,[1] M.P. Lilly,[1] M.A. Eriksson,[2] and M.S. Carroll[1]

[1]Sandia National Laboratories, Albuquerque, New Mexico 87123, USA
[2]University of Wisconsin-Madison, Madison, Wisconsin 53706, USA
[3]University of Oklahoma, Norman, Oklahoma 73019, USA

Abstract

We present measurements of silicon (Si) metal-oxide-semiconductor (MOS) nanostructures that are fabricated using a process that facilitates essentially arbitrary gate geometries. Stable Coulomb blockade behavior showing single period conductance oscillations that are consistent with a lithographically defined quantum dot is exhibited in several MOS quantum dots with an open lateral quantum dot geometry. Decreases in mobility and increases in charge defect densities (i.e. interface traps and fixed oxide charge) are measured for critical process steps, and we correlate low disorder behavior with a quantitative defect density. This work provides quantitative guidance that has not been previously established about defect densities and their role in gated Si quantum dots. These devices make use of a double-layer gate stack in which many regions, including the critical gate oxide, were fabricated in a fully-qualified complementary metal-oxide semiconductor facility.

## Introduction

Depletion-mode lateral quantum dots in GaAs/AlGaAs heterostructures have been used to make electrically controlled spin qubits.[1-6] Important features of the lateral geometry include a tunable-tunnel barrier opacity, a highly tunable electron occupation, and an open active region, providing the ability to tune the position of the quantum dots.[7-9] The lateral non-collinear geometry helps to maintain strong tunnel coupling to two different reservoirs as quantum dots decrease in size. This geometry has been used to enable one-electron occupation in single,[10] double,[11] and triple quantum dots.[12] Charge sensor integration in these structures is also facilitated by the lateral open geometry, and it assists in unambiguous identification of single electron occupation and spin measurement through spin-to-charge transduction. An additional aspect of the GaAs/AlGaAs material system is that all the nuclei have non-zero spin. Although the nuclear spins of both Ga and As have been used successfully to drive rotations in a singlet-triplet qubit,[13] and can be partially controlled,[14] nuclear spins are also an undesirable source of spin decoherence.

The existence of nuclear spin-free semiconductor materials such as $^{28}Si$ has motivated significant efforts to realize silicon quantum dot qubits.[15-17] Quantum dots have been formed in silicon by a wide variety of methods, including enhancement-mode silicon metal-oxide semiconductor (MOS) structures,[18-20] strained-Si/SiGe (sSi/SiGe) modulation-doped heterostructures,[21-25] collinear gated nanowires,[26] and atomically-patterned delta-layers.[27] The ability to form few electron singlet and triplet states has also been demonstrated with spin blockaded transport for both sSi/SiGe[28] and enhancement mode collinear MOS[29] double quantum dots.

Advantages of pursuing enhancement mode MOS structures include: (1) highly variable electron density in the two-dimensional electron gas (2DEG); (2) the possibility for outstanding charge stability;[30-32] (3) simpler integration with standard classical Si complementary MOS (CMOS) electronics for potential integration with high speed charge sensing circuitry;[33] and (4) potential hybrid integration with single donors.[34, 35] Investigation of enhancement mode devices is further motivated by the recognition that charge noise in modulation-doped heterostructures, while very small, is in fact associated with the modulation doping layer that is absent in an enhancement mode device.[36]

A common characteristic of silicon MOS enhancement mode quantum dots, to date, is that both the leads and the quantum dot(s) are collinear, with entrance and exit point contacts aligned across the device.[18-20, 29-32] This geometry encourages the dot to form in a small region with only modest flexibility available to tune the quantum dot position with respect to its leads, other neighboring quantum dots, and disorder in the system. In contrast, the gate designs most commonly used to form quantum dots in modulation doped heterostructures have a more open gate pattern and a non-collinear design.[10] In such open designs, in part because electrons are not confined to a narrow strip by etching or a narrow top gate, many properties of the device are more easily tuned *in situ*, including the position of the dot and the role of any particular gate, a feature that helps explain the ubiquity of this design. Demonstration of an open non-collinear lateral MOS quantum-dot geometry with single-period Coulomb blockade, consistent with the lithographic gate geometry and not dominated by the period of localized parasitic dots, would lead to a significant increase in flexibility of enhancement-mode MOS quantum dots and would represent a critical step towards advancing the MOS system along the same path used for spin qubits in GaAs/AlGaAs.

Disorder is produced by many factors, and the implications of the different forms of disorder for tunable quantum-dot behavior are not well characterized or understood. The MOS system is susceptible to fabrication-induced disorder, which can manifest as scattering and parasitic dot formation in transport, making it difficult to probe only the transport resulting from the lithographically gated quantum dot.[35, 37] One such factor is the introduction of charge defects. The density of such defects is a strong function of the device processing in MOS systems. Although there is an extreme wealth of knowledge about (a) silicon processing, (b) processing effects in classical room temperature field-effect transistors (FETs), and (c) some studies on collinear quantum-dot geometries,[37, 38] significant questions still remain about the implications the defects and disorder have for the viability of MOS quantum dots for coherent few electron spin physics. A necessary step towards addressing these questions is to examine quantum dot behavior with a fully characterized process flow.

In this paper we present (1) open-lateral MOS quantum dot transport that shows single-period Coulomb blockade over wide bias ranges; (2) single-period Coulomb blockade consistent with a quantum dot defined by the lithographic features and not dominated by disorder; (3) quantitative characterization of the process steps used to fabricate both the low disorder quantum dots as well as cases for which disordered quantum-dot behavior is strongly exhibited; and (4) numeric calculations of the corresponding magnitude of the disorder potential produced by these ranges of charge defects, which is compared to a modulation-doped GaAs/AlGaAs case. An important contribution of this paper, therefore, is to demonstrate single period, low disorder Coulomb blockade in an open MOS quantum dot geometry and identify a quantitative range of defect density and mobility for which this behavior occurs.

The paper is divided into three sections. In section I, we describe the device design and fabrication including the use of an initial MOS stack fabricated within a 0.35 µm CMOS production line that allows for flexible nanostructure post-processing. Defect characterization is described for critical nanostructure fabrication steps and includes mobility, interface trap density $D_{it}$, and effective fixed-oxide charge $Q_{fb}$. In section II, we report transport measurements of point contacts and quantum dots, including disorder free transport through devices with an open-lateral geometry. Section III discusses numerical calculations of both the disorder potential and the capacitance of a quantum dot using a three dimensional (3D) finite element commercial package that shows that the measured quantum dot capacitances are consistent with dots defined by the lithographic gates and not by a disorder potential.

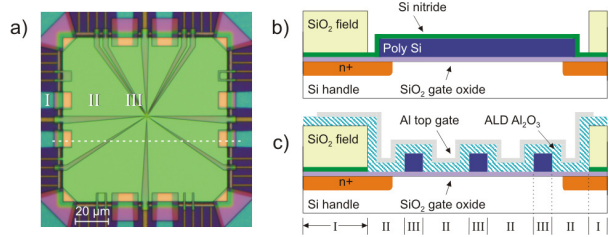

**Figure 1** (a) Optical image of an etch defined device before both secondary dielectric and global top gate deposition. The following features are identified: (I) $n+$ implanted ohmic contacts extending into a window etched into the deposited field oxide, (II) exposed gate oxide within the etched oxide window, and (III) etched polysilicon gates patterned with electron-beam lithography. (b) Cross-sectional schematic after processing in the CMOS facility. (c) Cross-sectional schematic of the completed layer structure corresponding to the line cut indicated by the dashed horizontal line in (a).

## I. Device Fabrication

The silicon quantum devices studied here are fabricated in two phases. A MOS gate stack with an exposed 100 µm x 100 µm degenerately doped polysilicon accumulation gate is fabricated using the same process line used for dedicated 350 nm CMOS circuit fabrication at Sandia National Laboratories. The gate stack is modified to allow a second phase of fabrication using electron-beam lithography (EBL), which is available only outside of the Si CMOS facility. These modifications include a window in the field oxide opened to expose the polysilicon for subsequent nanofabrication and peripheral $n+$ ohmic implants, providing electron reservoirs at the edge of the window, as seen in Figs. 1(a) and 1(b). Details of the phase I processing can be found in Appendix A.

**Table 1. Summary of mobility measurements**

| Condition | $T_{SiO2}$ [nm] | $T_{Al2O3}$ [nm] | $\mu_{peak}$ [cm²/Vs] | Start $\mu_{peak}$ [cm²/Vs] | $n_{peak}$ [cm⁻²] | $\mu(n=5 \times 10^{11}$ cm⁻²) [cm²/Vs] | Density Method | Oxidation Condition |
|---|---|---|---|---|---|---|---|---|
| Phase I only | 35 | 0 | 8 000 – 16 000 (wafer dependant) | 8 000 – 16 000 | ~$10^{12}$ | 5 500 - 10 000 | $C_{ox-nominal}$ | Various |
| Phase I / e-beam litho (EBL) | 10 | 0 | 2 000 | 10 000 | $3.5 \times 10^{12}$ | 650 | $C_{ox-nominal}$ | 10nm Ox. w/ DCE[a] |
| Phase I / e-beam litho (EBL) / forming gas | 10 | 0 | 6 600 | 10 000 | $2.4 \times 10^{12}$ | 3 500 | $C_{ox-nominal}$ | 10nm Ox. w/ DCE |
| Phase I / poly-silicon etch / Al[b] | 35 | 0 | 150 | 13 800 | $>4 \times 10^{12}$ | 30 | $C_{ox-nominal}$ | 35nm Ox. w/ DCE |
| Phase I / poly-silicon etch / Al / forming gas | 35 | 0 | 5 250 | 13 800 | $1.8 \times 10^{12}$ | 750 | Hall | 35nm Ox. w/ DCE |
| Phase I / poly-silicon etch / ALD / Al | 35 | 60 | 300 | 13 800 | $>4 \times 10^{12}$ | 50 | $C_{ox-nominal}$ | 35nm Ox. w/ DCE |
| Phase I / poly-silicon etch / ALD / Al / forming gas | 35 | 60 | 8 300 | 13 800 | $1 \times 10^{12}$ | 6 800 | Hall | 35nm Ox. w/ DCE |
| Phase I / ebeam Al | 35 | 0 | 4 950 | 14 250 | $2 \times 10^{12}$ | 2 500 | $C_{ox-nominal}$ | 35nm Ox. w/ DCE |
| Phase I / ebeam Al / forming gas | 35 | 0 | 8 700 | 14 250 | $1.2 \times 10^{12}$ | 5 700 | $C_{ox-nominal}$ | 35nm Ox. w/ DCE |
| Phase I / thermal Al | 35 | 0 | 8 000 | 8 000 | - | - | - | - |

[a]DCE – (Dichloroethane)
[b]Unless noted, Aluminum deposited via electron beam evaporation

The phase II fabrication of the nanostructure consists of EBL followed by a plasma etch to form the polysilicon depletion gates. A secondary dielectric, consisting of 60 nm of $Al_2O_3$, was added via atomic layer deposition (ALD), which was then annealed in forming gas. In some cases, reoxidation was done before the ALD step. Subsequent metallization for the top gate and its contacts were done to complete the structure, followed by a final forming-gas anneal. A schematic of the final gate stack can be seen in Fig. 1(c). Specific details of each step during phase II processing can also be found in Appendix A.

The disorder potential at the oxide-Si interface is influenced by a number of factors, including fixed defect charges within the oxide, interface trap charge, and surface roughness. Many processing steps in phase II can introduce damage resulting in additional charge within the structure and reducing the quality of the phase I material. Processing steps were characterized by examining changes in either mobility or charge defect density. High-frequency and quasi-static capacitance-voltage (C-V) were used to measure static oxide charge density and interface trap density.

Low-temperature mobilities ($T = 4.2$ K) were measured in test structures that simulated the process steps of the device fabrication. Initial (phase I only) peak mobilities were as high as 16 000 cm²/V s, with a wafer to wafer variability noted in Table 1. The experiments described in Table 1 include a mobility measurement of the phase I and mobility measurements of duplicate samples after each indicated phase II process step. The mobility measurements highlight that: (1) damage to the active region can occur despite an overlying protective layer of polysilicon, notably during EBL, which is a workhorse lithographic method of the quantum dot community; (2) mobilities can be very low (~200 cm²/Vs) after steps such as the polysilicon etch; (3) forming gas anneals are critical, and enable the recovery of a significant fraction of the mobility, but are not sufficient to recover the entire mobility; and (4) gate stacks most like the quantum dots with the lowest disorder in section II show a mobility of 8 300 cm²/Vs. The damage from the EBL is presumably a result of secondary photon emission (e.g., x-rays) due to the stopping of the impinging high-energy electrons,[40, 41] which penetrate the polysilicon and affect the underlying gate oxide and silicon. Although similar qualitative observations about process damage and forming-gas annealing exist in previous literature, few reports exist that describe the process effect on low-temperature mobility that is a commonly used figure of merit for low-temperature devices. In section II, correlation of these mobilities with the quantum-dot behavior leads to the observation of single-period Coulomb blockade in samples with both low and high mobility. While the higher mobility device does indeed display superior performance, it is clear that mobility alone is not a sufficient indicator of the ability to form tunnel barriers and quantum dots that exhibit single period Coulomb blockade consistent with a lithographically defined dot, in contrast to quantum dots formed by local trapping or a disorder potential.

**Table 2. Summary of Capacitance-Voltage Test Structures Simulating Device Process Flow**

| Condition | $T_{\text{SiO2-nominal}}$ [nm] | $T_{\text{Al2O3}}$ [nm] | $T_{\text{SiO2-meas}}$ [nm] | $D_{\text{it-lq}}$ [cm²/Vs] +/-1x10¹⁰ | $D_{\text{it-ay}}$ [cm²/Vs] | $V_{\text{fb}}$ [V] | $Q_{\text{fb}}$ [cm⁻²] +/-3.2x10¹⁰ |
|---|---|---|---|---|---|---|---|
| SiO₂[a] / Al | 35 | 0 | 35.3 | 2.39x10¹⁰ | 3.51x10¹⁰ | -0.79 | 6.58x10¹⁰ |
| SiO₂ / 15 nm Al2O3 / Al | 35 | 15 | 35 | 2.88x10¹⁰ | 2.71x10¹⁰ | -0.45 | -1.17x10¹¹ |
| SiO₂ / 30 nm Al2O3 / Al | 35 | 30 | 35 | 3.68x10¹⁰ | 4.19x10¹⁰ | -0.76 | 3.5x10¹⁰ |
| SiO₂ / poly-silicon etch / Al | 35 | 0 | 8.4 | - | - | - | - |
| SiO2 / Al / no FG anneal [39] | N/A | 0 | N/A | 3-6x10¹² | N/A | N/A | N/A |
| SiO₂[b] / Al | 10 | 0 | 10 | - | - | -0.72 | 8.73x10¹⁰ |
| SiO₂ / Al | 70 | 0 | 70 | 1.74x10¹⁰ | 2.55x10¹⁰ | -0.86 | 5.38x10¹⁰ |
| SiO₂ / poly-silicon etch / Al | 70 | 0 | 44.7 | 3.43x10¹⁰ | 3.4x10¹⁰ | -0.13 | -2.69x10¹¹ |
| SiO₂ / thermal Al | 35 | 0 | 35 | 3.32x10¹⁰ | 2.98x10¹⁰ | -1.25 | 2.71x10¹¹ |

[a] Oxides grown in oxygen ambient at 1atm with a temperature of 1000°C.
[b] 10 nm oxide grown in 10% oxygen partial pressure diluted with Ar at 1000°C

The relative change in interface trap density $D_{\text{it}}$ was also examined for several conditions similar to the device processing in an effort to find processes that minimize both $D_{\text{it}}$ and $Q_{\text{fb}}$ (Table 2). Details about the $D_{\text{it}}$ measurements are provided in Appendix B. Forming gas anneals are known to be critical in reducing $D_{\text{it}}$[40, 41] as well as recovering mobility. After forming gas annealing, little variation in the reduced $D_{\text{it}}$ is observed for the different process steps. Because $D_{\text{it}}$ and near interface traps are correlated with both $1/f$ noise and paramagnetically active defects, it is an important observation that the nanostructure fabrication steps do not introduce more of these harmful defects.

The flat-band voltage $V_{\text{fb}}$ was monitored for the different processing conditions, to examine the change in effective net charge in the oxide (Table 2). A shift in $V_{\text{fb}}$ can be associated with a change in the sum of positive and negative charge in the oxide stack. For unprocessed thermal gate oxides, it is known that a large positive fixed charge resides near the oxide-Si interface. $Q_{\text{fb}}$ is the charge calculated from the difference between the measured flat-band voltage shift and the theoretical ideal case with a charge-free oxide. The charge at the oxide-Si interface that accounts for the observed flat-band shift in each case investigated is reported in Table 2. Process damage can distribute either negative or positive charge throughout the oxide, with charge closer to the top metal gate contributing less of an offset than that nearer to the interface, due to screening from the conducting gate. $Q_{\text{fb}}$ is, therefore, an equivalent charge that assumes the charge within the oxide stack is near the interface. We note that mobility is sensitive to the total number of charge centers rather than the net charge.

$Q_{\text{fb}}$ is strongly dependent on the processing choices. The second dielectric thickness, the polysilicon etch,[42] and the method of Al deposition,[43] each impact the effective charge. The magnitudes of the charge densities range from ~5x10^10 to 3x10^11 cm^-2. In the cases where forming gas anneals are done, the largest contribution to disorder is the fixed charge, which will make a major contribution to the resulting electrostatic disorder potential. Calculations of electrostatic potentials resulting from this range of defects reveal RMS fluctuations in the potential comparable to a modulation-doped GaAs/AlGaAs case, as discussed below in section III. For cases in which no forming gas anneal is performed, literature reports suggest that the interface trap density can be greater than $10^{12}$ cm^-2/eV, well above the measured fixed charge density, and large enough that it would be predicted to produce very deep fluctuations in the potential near the Si interface (section III).

## II. Device Structure and Transport Measurements

The double-top-gate and open-lateral geometry discussed here provide flexibility in the location of the intended quantum dot relative to the leads and other neighboring nanostructures (e.g., point contacts and neighboring dots). The configuration used in this work is shown in Fig. 2(a). This structure emulates a gate configuration often used for modulation doped quantum dots and that has successfully been used to form spin qubits by other groups.[3-6] However, in the enhancement-mode MOS structure, a top Al gate overlays the polysilicon depletion gates, as shown in Figs. 2(b) & 2(c), inducing the conduction electrons in the leads and the dot. A 3D capacitance model, described later in this paper, addresses the capacitance matrix of this system and enables an understanding of the quantum dot size and location.

Transport measurements of two samples (A and B) were performed in a $^3$He refrigerator with a base temperature of 250 mK, and a third sample (C) was measured in a dilution refrigerator operating at a temperature of 100 mK. Conductance measurements were done using a lock-in amplifier with an AC excitation of 100 μV at 13 Hz, unless otherwise noted. Samples A and B are shown to have similar mobilities, 800 and 600 cm^2/Vs respectively, but sample B experienced processing known from the work presented in Sec. I to introduce more defects than that experienced by sample A. Sample C experienced processing that produces the lowest defect density of all three samples, and its field mobility is ~4500 cm^2/Vs. Additional information about the differences in sample fabrication can be found in Appendix A. The structure shown in Fig. 2(a) has a central region defined by depletion gates B - F and H that are operated to form a single oval quantum-dot region for devices A and B, and enables tunability through the use of many independent gates. Gates A and G form point contacts with B and F, respectively, and are intended for charge sensing (not discussed here). In spite of similar

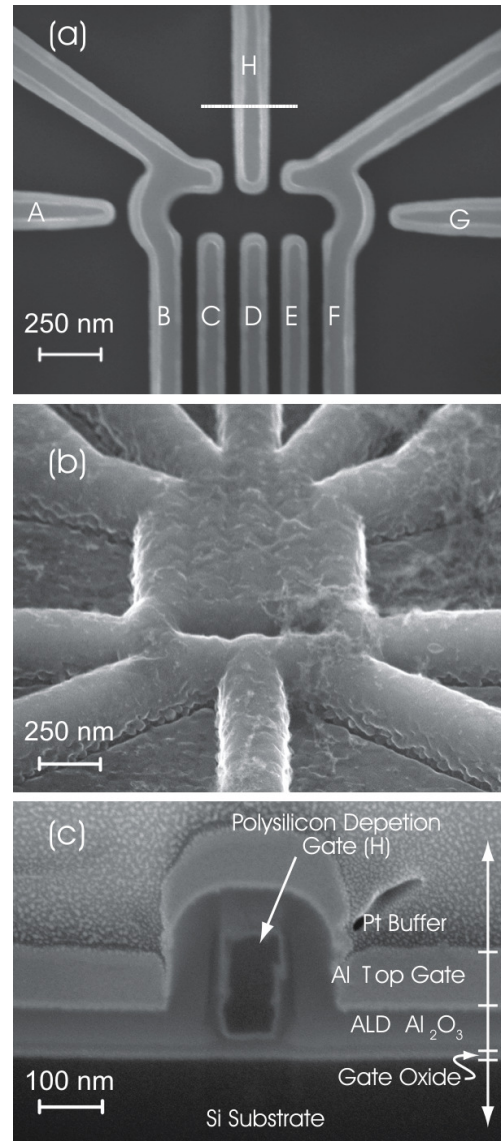

**Figure 2** (a) Scanning electron microscope (SEM) image of Si nanostructure without Al$_2$O$_3$ or Al top gate; (b) tilted SEM image of completed device with both Al$_2$O$_3$ and Al top gate; and (c) SEM cross section after focused ion-beam cut of a completed device through gate H at the position shown by the solid horizontal line in (a).

mobilities between samples A and B, the pinch-off-curves for point contacts formed in these two devices, shown in Fig. 3(a) and 3(b), are distinctly different: sample A shows a pinch-off curve with relatively low disorder. The pinch-off curves show a small number of resonances that are probably due to a combination of back-scattering,[44] a parasitic dot,[35] and quantum interference.[45] In contrast, sample B shows dramatic evidence of disorder in the form of resonances throughout pinch-off of various magnitudes and periods, likely indicating the sustained presence of one or more parasitic dots within the channel. The magnitude and frequency of scattering and trapping in such a defect rich channel make it exceptionally difficult to observe single-period Coulomb blockade, consistent with a lithographically formed dot, over any range of bias in sample B.

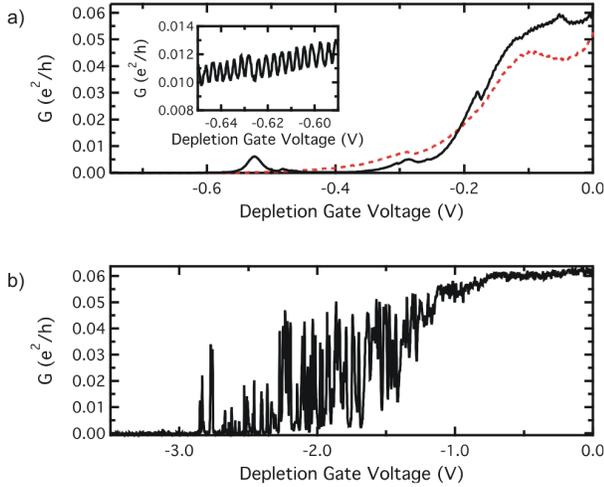

**Figure 3** (a) Transport through point contacts in sample A; Conduction is plotted as a depletion gate voltage is applied to both gates B and H (solid trace) and gates F and H (dotted trace) with a top-gate voltage of $V_{TG}$=25 V and an applied ac source-drain voltage of 100 $\mu$V. Inset: Transport through the gated quantum dot shown in Fig. 2a as a function of $V_D$ showing small amplitude Coulomb blockade oscillations $V_T$=25 V, $V_B$=$V_F$=0, $V_H$=-100 mV, $V_C$=$V_E$=-500 mV. (b) Transport through a point contact in sample B, which was fabricated without a forming gas anneal; Conduction is plotted as a depletion gate voltage is applied to both gates A and B with a top gate voltage of $V_{TG}$=8 V and an applied AC source-drain voltage of 100 $\mu$V. Instead of a monotonic decrease in conduction, disorder within the channel produces resonances through pinch off characteristic of the formation of quantum dots within the constriction.

The inset of Fig. 3(a) shows representative conductance measurements through the quantum dot. Periodic Coulomb blockade oscillations with similar peak heights were observed. Coulomb blockade oscillations were visible when sweeping any of the gates immediately adjacent to the quantum dot. In addition to periodic Coulomb oscillations, a noticeable decrease in overall conduction (both the maxima and the minima of the peaks) is observed as $V_D$ becomes increasingly negative in the inset of Fig. 3(a). This decrease is due to capacitive coupling between gate D and the entrance and exit tunnel barriers.

In addition to the modest difference in mobility, there are three important differences between samples A and B. First, the feature sizes of sample B are much larger, with gap sizes of 150 nm as opposed to 50 nm or smaller, resulting in much larger pinch-off voltages and a longer effective length for the point contact. Second, the polysilicon depletion gates of sample B were not subject to an additional oxidation after the pattern-etch. Finally, at no point during the processing of sample B was a forming-gas anneal performed. Because sample B was subjected to neither a second oxidation nor a forming-gas anneal, both structural and charge-based damage incurred during phase II processing remain as potential sources of disorder. The contrast in character of Figs. 3(a) and 3(b) suggests that additional anneals reduce the number of defects that can lead to scattering or parasitic dot formation, which correlates well with the observed reduction in oxide charge and interface traps (section I) and indicates that peak mobility alone is not a sufficient measure of the disorder that is relevant for quantum-dot devices.

We now discuss a device, sample C, that was processed using many of the optimizations described above. Measurement of test stacks with these additional changes led to peak mobilities higher than 8000 cm$^2$/Vs, approaching the baseline values obtained immediately after phase I processing.[46] The mobility measured in the field of sample C was $4500 \pm 500$ cm$^2$/Vs. While not as high as the test structure, this final device mobility is an improvement upon the mobilities of samples A and B.

Single quantum dots were observed in the sample C structure on both the left (gates B, C, D, and H) and right (gates D, E, F, and H) portions of this nanostructure, demonstrating the flexibility of the double layer gate design. Representative data is presented here on a single dot defined on the right side of the structure depicted in Fig. 2(a). The electronic configuration included setting $V_B$ = $V_C$ = +1.5 V to allow unrestricted conduction beneath gates B and C. Tunnel barriers were formed with gate pairs D and H, as well as F and H, while gate E was used as a plunger gate and was minimally coupled to either tunnel barrier. As shown in Fig. 4(a), stable Coulomb blockade was observed at and around the following voltage conditions: $V_{Top\ Gate}$ = +5 V, $V_D$ = -800 mV, $V_E$ = -1.25 V, $V_F$ = -2.3 V, and $V_H$ = -900 mV. Importantly, as a result of the processing modifications described above, the Si/SiO$_2$ interface and surrounding dielectric regions were clean enough to allow well-separated peaks with essentially zero conductance in between the Coulomb peaks. An envelope modulation in the Coulomb-blockade peak amplitude is observed in the Coulomb-blockade oscillations as a function of gate voltage. Coulomb-blockade amplitude envelope modulation has been observed in many other semiconductor quantum dot systems, and has been attributed to a variety of sources, including changes in coupling to the leads as orbitals are filled, quantum chaos, universal conductance fluctuations, and variations in tunnel barrier conductances due to either capacitive cross-talk or disorder.[7] Figure 4(b) shows measurements of Coulomb diamonds in this device. The differential conductance was measured using a lock-in technique with a source-drain voltage $V_{SD}$ modulated at 13 Hz with an excitation of 50 µV. The Coulomb diamonds indicate a charging energy $E_C \approx 1.1$ meV for this quantum dot.

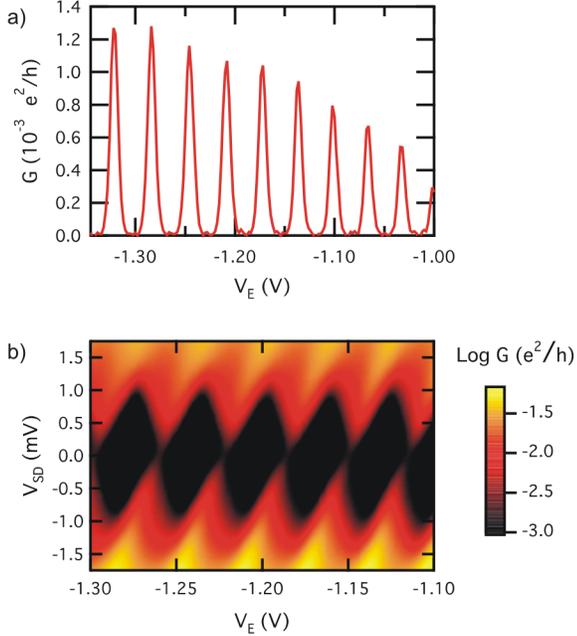

**Figure 4** (a) Conductance as a function of the voltage on gate E through the quantum dot, formed by gates D, E, F, and H shown in Fig. 2(a). The voltages on the remaining gates are: $V_{Top\ Gate}$ = +5V, $V_D$ = -800 mV, $V_F$ = -2.3 V, $V_H$ = -900 mV. (b) Conductance as a function of source-drain voltage and the voltage on gate E. Voltages on the remaining gates are: $V_{Top\ Gate}$ = +5V, $V_D$ = -800 mV, $V_F$ = -2.1 V, $V_H$ = -900 mV.

## III. Discussion and Analysis

To establish whether the observed quantum dot behavior was consistent with confinement due to the lithographic gates, in contrast with smaller disorder dots, capacitances for the quantum dot structure were calculated numerically and compared to experimental results. A combination of commercial simulation packages were used to calculate the capacitance network of the 3D structure. The initial step was to solve for the potential and charge distribution in the quantum dot region using a technology computer assisted design (TCAD) package, Davinci, results of which are shown within Fig. 5(a) for the same gate bias as was used for the transport measurement shown in the inset of Fig. 3(a). A contour of constant electron density 1.5 x $10^{11}$ cm$^{-2}$ is shown in Fig. 5(a) and was used to define the metallic edge of the quantum dot. An electron density of approximately 1.5 x $10^{11}$ cm$^{-2}$ has been observed to show a metal-insulator transition at ~100 mK for MOSFETs with mobilities of ~10 000 cm²/Vs.[47] The edge of this density contour was used to define the geometry of a metal sheet, shown in Fig. 5(b), for a 3D finite elements calculation of the capacitive coupling to the depletion gates. The 3D structure was constructed from the 2D computer assisted design (CAD) mask levels and nominal layer thicknesses, Figs. 5(c), 5(d), and 5(e). Table 3 shows the capacitances calculated with CFD-ACE+, a multiphysics software package from ESI Group. These capacitances are compared to experimental values obtained from device A, measured by monitoring the period of Coulomb blockade oscillations as each specific depletion gate voltage is changed. The modeled capacitances are in good agreement with the experimental values, both in magnitude and trend, consistent with the interpretation that the measured quantum dots are defined by the lithographic features of the tunable gated structure rather than by a disorder potential.

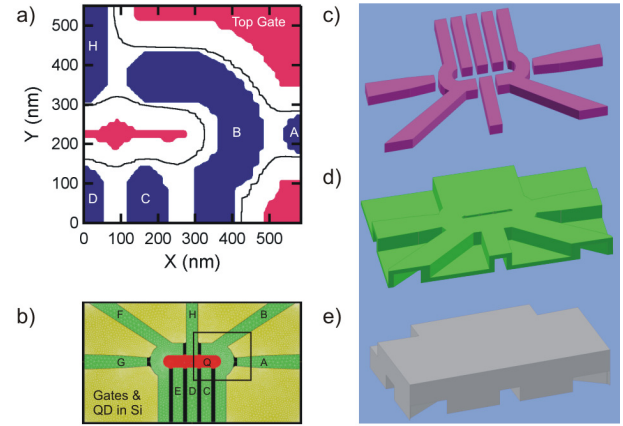

**Figure 5** (a) Results of an electrostatic simulation of a quantum dot structure similar to that shown in Fig. 2(a). Blue regions in the figure represent depletion gates, magenta regions correspond to areas where the metallic top gate comes within a vertical distance of 95nm of the Si-SiO$_2$ interface, as noted. The locations of the 2DEG leads and the quantum dot are shown using a black contour marking the conduction edge. b) Gate labels and representative meshing for 3D capacitance modeling. The area indicated by the black box corresponds to the portion of the device modeled in (a). (c), (d), and (e) 3D rendering of the doped polysilicon depletion gate layer, the dielectric layer, and the metallic top gate layer respectively used in the 3D capacitance modeling. The results of which are shown in Table 3.

Defect charges at or near the oxide-Si interface can contribute to disorder by forming random fluctuation potentials deep enough to trap or scatter electrons. Having measured a range of defect densities in the oxide, the dependence of the average magnitude of the disorder potential can be calculated for this range to offer insight about whether the higher range of measured charge densities could contribute sufficiently to explain the disorder in sample A, in contrast with other possible contributions to the disorder such as surface roughness. In particular, over some bias ranges the peaks in the disordered barriers in Fig. 3(b) are equally spaced, indicative of many-electron parasitic dot formation, and inconsistent with trapping from single isolated positive charges in the oxide. The latter would be expected to bind no more than a few electrons, in analogy with an ionized donor in silicon with a confining potential of ~40-50 meV.

**Table 3. Calculated and Measured Capacitances from Quantum Dot Sample A**

| Gate | Modeled Capacitance | Measured Capacitance |
|------|---------------------|----------------------|
| B-Q | 40.2 aF | 48.1 aF |
| C-Q | 16.4 aF | 14.7 aF |
| D-Q | 16.9 aF | 11.2 aF |
| E-Q | 16.4 aF | 14.1 aF |
| F-Q | 40.2 aF | 48.1 aF |
| H-Q | 16.7 aF | 20.8 aF |

Numerical simulations of the potential produced by a random distribution of bare positive charges located within the oxide and 2 nm from a Si-oxide interface are shown for values of $1 \times 10^{10}$ cm$^{-2}$, $1 \times 10^{11}$ cm$^{-2}$, and $1 \times 10^{12}$ cm$^{-2}$ in Figs. 6(a), 6(b), and 6(c), assuming the centroid of the electron distribution is ~5 nm from the interface. The case where the net charge is zero through the introduction of both positive and negative charge densities of magnitude $5 \times 10^{10}$ cm$^{-2}$ (a total to $1 \times 10^{11}$ cm$^{-2}$ scattering centers) is also shown in Fig. 6(d), demonstrating the effect of compensating negative charge on the disorder potential. An image charge solution was used to calculate the potential in the silicon due to a charge in the oxide,[48] using a dielectric constant of 11.9 for silicon and 3.9 for silicon dioxide. The standard deviation of the potentials summed over 20 different ensembles was calculated to be 4.3, 13.6, 42.0, and 13.5 meV for $10^{10}$, $10^{11}$, $10^{12}$, and $-5 \times 10^{10}$ cm$^{-2}$ $+ 5 \times 10^{10}$ cm$^{-2}$ charge densities with maximum well depths of 40.7, 201, 1630, and 48.0 meV, respectively, where the variance in charge position leads to clusters of charge with significantly deeper wells. The standard deviation in a high mobility modulation-doped GaAs/AlGaAs structure with $2.16 \times 10^{12}$ cm$^{-2}$ ionized donors at a distance of 28 nm, for comparison, was calculated by Nixon and Davies[49] to be 18 mV, which falls within the range of $10^{11}$ and $10^{12}$ cm$^{-2}$ fixed charge at the Si/SiO$_2$ interface.

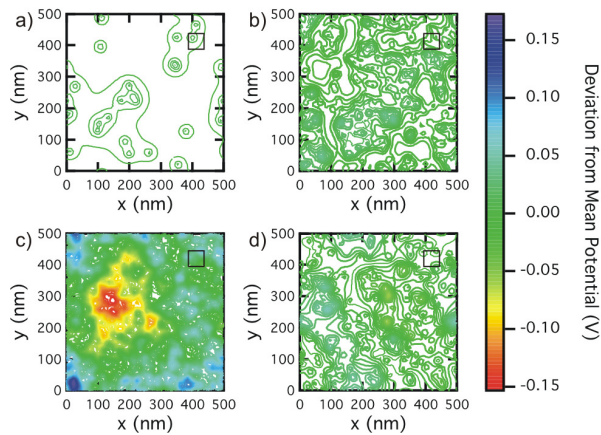

**Figure 6** A simulation of bare potential with random distribution of charges located within the oxide, 2nm from the Si-SiO$_2$ interface with an charge density of (a) $1 \times 10^{10}$ charges/cm$^2$, (b) $1 \times 10^{10}$ charges/cm$^2$, (c) $1 \times 10^{11}$ charges/cm$^2$, and (d) +5 x $10^{10}$ charges/cm$^2$ combined with -5 x $10^{10}$ charges/cm$^2$ for a net zero charge at the interface. Contour spacing in each plot is 6mV. Boxes represent a guide for an area of approximately 50 nm x 50 nm representative of dimensions approximately the size of the point contact dimension.

Summary

Single-period Coulomb blockade, consistent with the charging energy of a lithographically formed quantum dot, is observed in two Si structures that have an open lateral MOS enhancement-mode gate geometry. The open-lateral gate geometry allows significant tuning of the confining potential and emulates structures in GaAs/AlGaAs and sSi/SiGe that have been used to achieve few electron double quantum dots and probe coherent few electron-spin properties. Stable dot formation in several different configurations (i.e., left, right, or both sides of a double-dot structure) is demonstrated showing the flexible tuning of the position of the dots in the open-lateral geometry. Achievement of low disorder quantum dots in an open-lateral enhancement-mode MOS geometry is motivated by potential benefits such as more tunable electron density, integration for enhanced read out, reduced background dopants, and potential hybrid donor quantum-dot coupling.

Disorder due to defects introduced during processing is found to be a critical hurdle to achieving quantum dots that are not plagued by scattering and parasitic dot formation. Characterization of each process flow step's effect on mobility and charging of the oxides through high and low frequency capacitance-voltage measurements were described. The primary change in process-induced charging of the oxides is in fixed charge in the oxide, while the interface trap density can be maintained near the same values as the starting gate oxide-silicon interface. Improved processing leads to a transition from a device that is plagued by disorder to one that shows clean quantum dot behavior over wide bias ranges. The open geometry is also used to form tunable dots in several configurations, consistent with using different lithographic gates to define the dot size rather than being dominated by disorder potentials. Single period Coulomb blockade behavior is achieved in process flows that are characterized to have mobilities that range from 500 to 4000 cm$^2$/Vs, $D_{it} \sim 3 \times 10^{10}$ cm$^{-2}$ eV$^{-1}$, and $Q_{fb} \sim 3.5 \times 10^{11}$ cm$^{-2}$, where $Q_{fb}$ is a net charge density of positive and negative charges combined. These charge densities correspond to an RMS electrostatic potential near the Si interface that is comparable in order of magnitude to a representative GaAs/AlGaAs modulation doped structures.

Acknowledgements

The authors are grateful for sample preparation done by D. Tibbets. The authors would also like to thank N. Bishop and E. Bielejec for reviewing the manuscript and helpful discussions. This work was performed, in part, at the Center for Integrated Nanotechnologies, a U.S. DOE, Office of Basic Energy Sciences user facility. The work at both Sandia National Laboratories and the University of Wisconsin was supported by the Sandia National Laboratories Directed Research and Development Program. Sandia National Laboratories is a multi-program laboratory operated by Sandia Corporation, a Lockheed-Martin Co., for the U. S. Department of Energy under Contract No. DE-AC04-94AL85000.

Appendix A:

Phase I Fabrication Steps:

1) Source-Drain Implant: $2 \times 10^{15}$ cm$^{-2}$ As implanted at an energy of 50 keV
2) Gate Oxidation: 35 nm SiO$_2$ grown @ 900°C for 150 min
3) Silicon Deposition: 200 nm nominally undoped amorphous silicon (a-Si)
4) a-Si Doping: $5 \times 10^{15}$ cm$^{-2}$ As implanted at an energy of 35 keV
5) a-Si Patterning, Etch: CF$_4$ based plasma etch through ~80% of the layer followed by a high selectivity HBr etch to stop on the SiO$_2$ gate oxide
6) a-Si Oxidation, Crystallization, and Activation: 10nm SiO$_2$ grown for 13 min @ 900°C, 30 min N$_2$ anneal @ 900°C
7) Field Oxide Deposition: 300-500 nm SiO$_2$ deposited via high density plasma chemical vapor deposition
8) Via Formation: CF$_4$ based plasma etch
9) Via, Bond Pad Metallization: 200 Å Ti, 500 Å Ti/N, 5000 Å W conformally deposited via chemical vapor deposition (CVD)
10) Oxide Window Opening: CF$_4$ based plasma etch

Phase II Fabrication Steps for Samples A & B:

1) e-beam Lithography: 100 keV electron beam with 400 pA beam current, and areal dose of 190 µC/cm$^2$ on negative e-beam resist (NEB)
2) Nitride/Oxide Etch: 440 second, CHF$_3$ based, 350 Watts, inductive coupled plasma reactive ion etch (ICP-RIE)
3) Depletion Gate Formation, polysilicon Etch: 170 second, HBr, 300 Watt ICP-RIE with exposed and developed NEB as etch mask
4) NEB Removal: Solvent clean, and 20 min. oxygen ash
5) Pre-Oxidation Clean: 15 min W etch in 60°C H$_2$O$_2$, 5 min Nitride etch in 165°C H$_3$PO$_4$, RCA Clean
6) Thermal Oxidation (Device A Only): 24 min thermal oxidation @ 900°C
7) Secondary Oxide Deposition: 200°C atomic layer deposition (ALD) of 60 nm of Al$_2$O$_3$ using trimethylaluminum and water (H$_2$O) precursor chemistry
8) ALD Via Etch: Patterned HF via etch to ohmic and polysilicon contacts through ALD Al$_2$O$_3$
9) Top Gate, Bonding Pad Metallization: 1000 Å Al deposited via e-beam evaporation
10) Forming Gas Anneal (Device A Only): 30 minutes @ 400°C (10% H$_2$)

Phase II Fabrication Steps for Sample C:

1)      e-beam Lithography:  100 keV electron beam with 400 pA beam current, and areal dose of 190 $\mu C/cm^2$ on negative e-beam resist (NEB)
2)      Nitride/Oxide Etch:  440 second, $CHF_3$ based, 350 Watts, inductive coupled plasma reactive ion etch (ICP-RIE)
3)      NEB Removal:  Solvent Clean
4)      Depletion Gate Formation, polysilicon Etch:  170 second, HBr, 300 Watt ICP-RIE with exposed and developed NEB as etch mask
5)      Pre-Oxidation Clean:  15 min W etch in 60°C $H_2O_2$, 5 min Nitride etch in 165°C $H_3PO_4$, RCA Clean
6)      Thermal Oxidation:  24 min thermal oxidation @ 900°C
7)      Secondary Oxide Deposition:  200°C atomic layer deposition (ALD) of 60 nm of $Al_2O_3$ using trimethylaluminum and water ($H_2O$) precursor chemistry
8)      Forming Gas Anneal:  30 minutes @ 450°C (10% $H_2$)
9)      Top Gate Metallization:  1000 Å Al deposited globally via e-beam evaporation
10)     Top Gate Formation:  Top Gate patterned and formed using MIF 319 developer as an Al etchant
11)     ALD Via Etch:  Patterned HF via etch to ohmic and polysilicon contacts through ALD $Al_2O_3$
12)     Bonding Pad Metallization:  Bond pads patterned, 1000 Å Al deposited via e-beam evaporation
13)     Final Forming Gas Anneal:  30 minutes @ 400°C (10% $H_2$)

<u>Appendix B</u>

Capacitance-voltage (C-V) measurements were performed on experimentally simulated gate stacks for this work. Phase I material was not used because of parasitic capacitances and the high resistance of the substrates, which is problematic for high frequency C-V measurements. The starting substrates used in this study were *p*-type epitaxial Si on $p^+$ substrates. Thermal gate oxides were grown at 1000°C and annealed in nitrogen at 1000°C for 15 min.

A near mid-gap interface trap density, $D_{it-lo}$, an average interface trap density, $D_{it-av}$, and $Q_{fb}$ are reported in Table 2 for the different cases examined. $D_{it}$ is reported at the lowest capacitance ($C_{lo}$) during the low frequency sweep as $D_{it-lo}$. $D_{it-av}$ is an average value of $D_{it}$ between the Fermi level and weak inversion. The uncertainty of $D_{it}$ for values around mid-gap as indicated in the table are dominated by a confidence of ~2 pF in the total capacitance value. The uncertainty for $D_{it-lo}$ is estimated to be $1x10^{10}$ cm$^{-2}$eV$^{-1}$. The uncertainty in $Q_{fb}$ arises from an uncertainty in the work function difference between the metal and the silicon ohmic contact, which is dependent on deposition conditions. For this work, the work function difference for e-beam deposited Al was measured to be 4.43 V determined from the flat band voltage dependence on oxide thickness and the uncertainty in that work function leads to a $3.2x10^{10}$ cm$^{-2}$ uncertainty in $Q_{fb}$.[50]

We note that in the etch experiment some of the oxide was lost during the polysilicon etch leading to a thinner oxide, as noted in Table 2. The experiment was done for both 35 nm and 70 nm thicknesses. Within the uncertainty of the measurement, $D_{it}$ both before and after the etches with a subsequent forming gas anneal is indistinguishable for the 70 nm case. Leakage through the thinned 35 nm oxide led to inaccurate C-V curves for that case. $D_{it-lo}$ of samples with 15, and 30 nm of ALD $Al_2O_3$ have a $D_{it-lo}$ of $2.9 x 10^{10}$ and $3.7 x 10^{10}$, respectively. The increase in $D_{it-lo}$ is within the uncertainty of the measurement, and therefore is not considered meaningful.